# Nuclear Spin Driven Quantum Tunneling of Magnetization in a New Lanthanide Single-Molecule Magnet: Bis(phthalocyaninato)holmium anion

Naoto Ishikawa,*,† Miki Sugita,† and Wolfgang Wernsdorfer*,‡

*Department of Chemistry, Tokyo Institute of Technology, O-okayama, Meguro-ku, Tokyo 152-8551, Japan, and Laboratoire Louis Néel, CNRS, BP 166, 25 Avenue des Martyrs, 38042 Grenoble Cedex 9, France*

RECEIVED DATE    E-mail: ishikawa@chem.titech.ac.jp, wernsdor@grenoble.cnrs.fr

Observation of staircase-like magnetization hysteresis loops of the "single-molecule magnets" (SMMs)[1,2,3] has generated much attention to the quantum nature of these compounds. At each step relaxation of magnetization of an SMM occurs through a quantum-tunneling path.[4,5] The discovery of this phenomenon, also evidenced by temperature independent relaxation[6-11] and quantum phase interference[12] has led to the idea of using SMMs for quantum computing.[13]

Most SMMs are composed of a high-spin polynuclear transition-metal complex with a high axial magnetic anisotropy. Another promising approach to construct SMMs is to use high-spin lanthanide ions as the magnetic center.[14] Alternating current (ac) magnetic susceptibility measurements above 2 K for the six heavy lanthanide complexes with phthalocyanines, $[(Pc)_2Ln]^-$ (Pc denotes phthalocyaninato, Ln = Tb, Dy, Ho, Er, Tm or Yb) showed that the terbium and dysprosium complexes having large axial magnetic anisotropies behave as SMMs.[14,15] The magnetic anisotropy change from easy-axial to easy-planar type occurs between the Ho and Er complexes in the series of the above six Pc double-decker complexes.[16,17] It has therefore been of great interest to show whether the Ho complex with an axial anisotropy can show SMM behavior at low temperatures.

In this communication, we report the first magnetization-versus-field measurement in the subkelvin range on $[(Pc)_2Ho]^-$ doped in diamagnetic $[(Pc)_2Y]^-\cdot TBA^+$ ($TBA^+$ = tetrabutylammonium cation). We show that the Ho complex exhibit not only hysteresis loops, but also resonant quantum tunneling of magnetization (QTM), which is a characteristic feature of SMMs.

**Scheme 1.** $[Pc_2Ho]^-$.

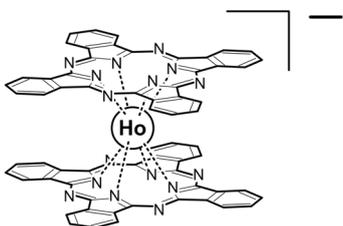

The compounds were prepared as reported in literature[18,19] with certain modifications.[16] The doped single crystalline samples were prepared by recrystallization from an acetone solution of the two compounds $[Pc_2Ho]^-\cdot TBA^+$ and $[Pc_2Y]^-\cdot TBA^+$, the masses of which were measured by an analytical balance. The crystal structure of the former compound has been reported by Koike, et.al.[20] Because of the strong resemblance between the ionic radius of $Ho^{III}$ (1.155 Å) and $Y^{III}$ (1.159 Å),[21] the two chemically analogous compounds are highly expected to have isostructural crystal structures. All measurements were performed using the micro-SQUID technique.[22] The field was aligned parallel to the easy axis of magnetization, using the transverse field method,[23] which was found to be parallel to the crystallographic c-direction.

Figure 1 shows magnetization vs. field measurements for a single crystal with the ratio of [Ho]/[Y]=1/49. Hysteresis is observed below 0.5 K and a step structure emerges. The hysteresis loops at 0.04 K (Figure 1-b) show a clear staircase-like structure indicating the occurrence of QTM. The steps are positioned equidistantly along the magnetic field axis at $\mu_0 H_n = n \times 23.5$ mT ($n = 0, 1, 2, 3…$). These steps were strongly broadened for the undiluted sample.

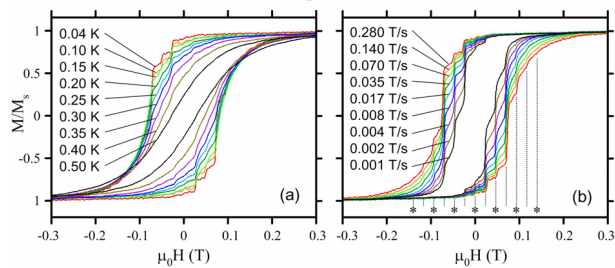

**Figure 1.** (a) Hysteresis loops at several temperatures for a single crystal of $[(Pc)_2Ho_{0.02}Y_{0.98}]^-\cdot TBA^+$ measured at the field scan rate of 0.28 T/s. (b) Hysteresis loops at 0.04 K measured at several field scan rates.

The ligand field (LF) splitting of the $[(Pc)_2Ln]^-$ complexes has been determined previously by "simultaneous optimization" of static magnetic susceptibility and $^1$H-NMR paramagnetic shift data.[16,17] The lowest substate in the $^5I_8$ ground multiplet of $[(Pc)_2Ho]^-$ was shown to be the $J_z = \pm 5$ doublet.[16] Under the magnetic field of the strength in the present paper, there is no crossing between sublevels of different $|J_z|$ values because of the large zero-field splitting due to the LF term. This situation clearly contrasts with the transition-metal-cluster SMMs, in which energy of substates with different $|S_z|$ values can coincide to give rise to a QTM process under an magnetic field below 1T because energy separations between substates are of the order of 1 to 10 cm$^{-1}$. The QTM observed for $[(Pc)_2Ho]^-$ cannot be attributed to the same mechanism as in the transition-metal-cluster SMMs.

Holmium has a nucleus with $I = 7/2$ spin in a natural abundance of 100 %. Each sublevel of the $J = 8$ ground multiplet is split into an octet by the hyperfine interaction between the $(4f)^{10}$ system and the nucleus. We performed exact numerical diagonalization of a $(2J+1)(2I+1) \times (2J+1)(2I+1)$ matrix including the hyperfine interaction $A_{hf}\mathbf{J}\cdot\mathbf{I}$ and the LF terms $A_k^0\langle r^k\rangle O_k^0$ ($k = 2, 4$ and $6$) with the LF parameters $A_k^0\langle r^k\rangle$ determined previously.[16]

Figure 2-a shows the Zeeman diagram for the sixteen $|J_z\rangle|I_z\rangle$ states created from the combinations of the $J_z = \pm 5$ lowest doublet and $I = 7/2$ octet. The level intersections are seen at 15

magnetic field positions. All observed step positions are reproduced with $A_{hf} = 0.0276$ cm$^{-1}$.

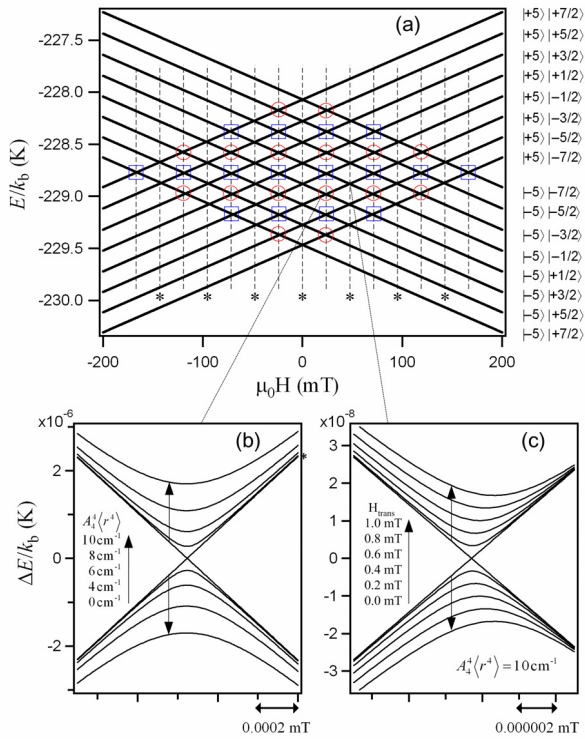

**Figure 2.** (a) Zeeman energy diagrams as a function of longitudinal magnetic field for the lowest $J_z = \pm 5$ substates with $I = 7/2$ nucleus calculated with the LF parameters determined in ref. 16 and $A_{hf} = 0.0276$ cm$^{-1}$. Squares, circles, and asterisks are explained in the text. (b) Plot for the intersection of $|+5\rangle|-3/2\rangle$ and $|-5\rangle|+1/2\rangle$ with several values for $A_4^4\langle r^4\rangle$. (c) The intersection of $|+5\rangle|-3/2\rangle$ and $|-5\rangle|-1/2\rangle$ as a function of transverse magnetic field H$_{trans}$.

For QTM to occur at a level intersection, the two states must be coupled leading to an avoided level crossing. Such coupling within the $J_z = \pm 5$ manifold is achieved by off-diagonal LF terms, such as $A_4^4\langle r^4\rangle O_4^4$ and $A_6^4\langle r^6\rangle O_6^4$, and the hyperfine-coupling term $A_{hf}\mathbf{I}\cdot\mathbf{J}$. The former terms couple $|J_z\rangle|I_z\rangle$ and $|J_z+4\rangle|I_z\rangle$, while the latter $|J_z\rangle|I_z\rangle$ and $|J_z+1\rangle|I_z-1\rangle$.

The 64 intersections can be grouped into three types. At the intersections indicated by circles, non-zero $O_4^4$ or $O_6^4$ terms yield avoided level crossings as shown in Figure 2-b. At the positions marked by squares, avoided crossings can occur with a non-zero $O_2^2$ term, which is possible when the four-fold symmetry is lost. For the rest of crossing points, which are labeled by asterisks, the presence of a transverse magnetic field is required to give avoided crossings as shown in Figure 2-c. Thus the steps marked by asterisks in Figure 1-b are assigned to this type.

Small steps which cannot be categorized in any of the three types above are observed at $\mu_0 H = \pm 10$ mT as shown in Figure 3-a. The height of these steps is significantly increased for higher Ho concentrations (Figure 3-b). They are most probably due to the two-body tunnel transitions such as the spin-spin cross-relaxation (SSCR)[24] mediated by magnetic dipolar interactions between Ho complexes, which have been recently observed for a scheelite-structured compound LiYF$_4$ doped with Ho ions [25] and for Mn$_4$ SMMs.[26]

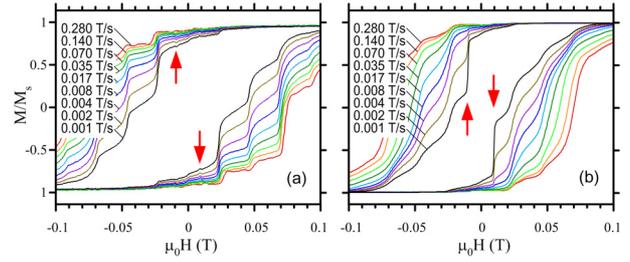

**Figure 3.** (a) Enlargement of hysteresis loops in Figure 1-b for a single crystal of [(Pc)$_2$Ho$_{0.02}$Y$_{0.98}$]$^-$·TBA$^+$. (b) Hysteresis loops at the same conditions for [(Pc)$_2$Ho$_{0.2}$Y$_{0.8}$]$^-$·TBA$^+$. The height of the steps indicated by the arrows is increased.

In conclusion, we have shown QTM in [(Pc)$_2$Ho]$^-$, a new lanthanide single-ion SMM. The quantum process is due to resonant quantum tunneling between entangled states of the electronic and nuclear spin systems, which is an essentially different mechanism from those of the transition-metal-cluster SMMs.

* To whom correspondence should be addressed.
† Tokyo Institute of Technology, Japan
‡ Laboratoire Louis Néel, France

(1) Christou, G.; Gatteschi, D.; Hendrichson, D. N.; Sessoli, R. MRS Bull. **2000**, *25* (11), 66.
(2) (a) Sessoli, R.; Tsai, H.-L.; Schake, A. R.; Wang, S.; Vincent, J. B.; Folting, K.; Gatteschi, D.; Christou, G.; Hendrichson, D. N. *J. Am. Chem. Soc.* **1993**, *115*, 1804. (b) Sessoli, R.; Gatteschi, D.; Caneschi, A.; Novak, M. A. *Nature* **1993**, *365*, 141.
(3) Aubin, S. M. J.; Wemple, M. W.; Adams, D. M.; Tsai, H.-L.; Christou, G.; Hendrichson, D. N. *J. Am. Chem. Soc.* **1996**, *118*, 7746.
(4) Friedman, J. R.; Sarachik, M. P. *Phys. Rev. Lett.* **1996**, *76*, 3830.
(5) Thomas, L.; Lionti, F.; Ballou, R.; Gatteschi, D.; Sessoli, R.; Barbara, B. *Nature* **1996**, *383*, 145.
(6) Aubin, S. M. J.; Dilley, N. R.; Wemple, Maple, M. B.; Christou, G.; Hendrichson, D. N. *J. Am. Chem. Soc.* **1998**, *120*, 839.
(7) Aubin, S. M. J.; Dilley, N. R.; Pardi, L.; Krzystek, J.; Wemple, M. W.; Brunel, L. C.; Maple, M. B.; Christou, G.; Hendrichson, D. N. *J. Am. Chem. Soc.* **1998**, *120*, 4991.
(8) Sangregorio, C.; Ohm, T.; Paulsen, C.; Sessoli, R.; Gatteschi, D. *Phys. Rev. Lett.* **1997**, *78*, 4645.
(9) Brechin, E. K.; Boskovic, C.; Wernsdorfer, W.; Yoo, J.; Yamaguchi, A.; SaVudo, E. C.; Concolino, T. R.; Rheingold, A. L.; Ishimoto, H.; Hendrickson, D. N.; Christou, G. *J. Am. Chem. Soc.* **2002**, *124*, 9710.
(10) (a) Soler, M.; Wernsdorfer, W.; Folting, K.; Pink, M.; Christou, G. *J. Am. Chem. Soc.*, **2004** *126*, 2156; (b) Soler, M.; Rumberger, E.; Folting, K.; Hendrickson, D. N.; Christou, G. *Polyhedron* **2001**, *20*, 1365.
(11) Andres, H.; Basler, R.; Blake, A. J.; Cadiou, C.; Chaboussant, G.; Grant, C. M.; GJdel, H. -U.; Murrie, M.; Parsons, S.; Paulsen, C.; Semadini, F.; Villar, V.; Wernsdorfer, W.; Winpenny, R. E. P. *Chem. Eur. J.* **2002**, *8*, 4867.
(12) Wernsdorfer, W.; Sessoli, R. *Science* **1999**, *284*, 133.
(13) Leuenberger, M. N.; Loss, D. *Nature* **2001**, *410*, 789.
(14) Ishikawa, N.; Sugita, M.; Ishikawa, T.; Koshihara, S.; Kaizu, Y. *J. Am. Chem. Soc.* **2003**, *125*, 8694.
(15) Ishikawa, N.; Sugita, M.;. Ishikawa, T.; Koshihara, S.; Kaizu, Y. *J. Phys. Chem. B* **2004**, *108*, 11265.
(16) Ishikawa, N.; Sugita, M.; Okubo, T.; Tanaka, N.; Iino, T.; Kaizu, Y. *Inorg. Chem.* **2003**, *42*, 2440.
(17) Ishikawa, N. *J. Phys. Chem. A* **2003**, *106*, 9543.
(18) De Cian, A.; Moussavi, M.; Fischer, J.; Weiss, R. *Inorg. Chem.* **1985**, *24*, 3162.
(19) Konami, H.; Hatano, M.; Tajiri, A. *Chem. Phys. Lett.* **1989**, *160*, 163.
(20) Koike, N.; Uekusa, H.; Ohashi, Y.; Harnoode, C.; Kitamura, F.; Ohsaka, T.; Tokuda, K. *Inorg. Chem.* **1996**, *35*, 5798.
(21) Shannon, R. D. *Acta Crystallogr., Sect. A* **1976**, 32, 751.
(22) Wernsdorfer, W. *Adv. Chem. Phys.* **2001**, *118*, 99.
(23) Wernsdorfer, W.; Chakov, N. E.; Christou, G. *Phys. Rev. B* **2004**, *70*, 132413.
(24) Bloembergen N. et al., *Phys. Rev.* (Utrecht) **1959**, *114*, 445.
(25) Giraud, R.; Wernsdorfer, W.; Tkachuk, A. M.; Mailly, D.; Barbara, B. *Phys. Rev. Lett.* **2001**, *87*, 057203.
(26) Wernsdorfer, W.; Bhaduri, S.; Tiron, R.; Hendrickson, D.; Christou, G. *Phys. Rev. Lett.* **2002**, *89*, 197201.